\documentclass[aps,pre,reprint,footinbib,superscriptaddress,citeautoscript]{revtex4-1}

\usepackage{amssymb}
\usepackage{amsmath}
\usepackage{amsfonts}
\usepackage{graphicx}
\usepackage{dcolumn}
\usepackage{bm}
\usepackage[english]{babel}
\usepackage[utf8]{inputenc}
\usepackage{color}
\usepackage{booktabs}
\usepackage{times}

\graphicspath{{.}{./Figures/}}

\begin{document}

\title{Fame and Obsolescence: Disentangling growth and ageing dynamics of patent
citations}

\author{K.~W. Higham}
\affiliation{Te P{\=u}naha Matatini, School of Chemical and Physical Sciences,
Victoria University of Wellington, PO Box 600, Wellington 6140, New Zealand}

\author{M. Governale}
\affiliation{Te P{\=u}naha Matatini, School of Chemical and Physical Sciences,
Victoria University of Wellington, PO Box 600, Wellington 6140, New Zealand}

\author{A.~B. Jaffe}
\affiliation{Te P{\=u}naha Matatini, Motu Economic and Public Policy Research,
PO Box 24390, Wellington 6142, New Zealand}

\author{U. Z\"ulicke}
\affiliation{Te P{\=u}naha Matatini, School of Chemical and Physical Sciences,
Victoria University of Wellington, PO Box 600, Wellington 6140, New Zealand}

\date{\today}

\begin{abstract}

We present an analysis of citations accrued over time by patents granted by the
United States Patent and Trademark Office in 1998. In contrast to previous
studies, a disaggregation by technology category is performed, and exogenously
caused citation-number growth is controlled for. Our approach reveals an
intrinsic citation rate that clearly separates into an -- in the long run,
exponentially time-dependent -- ageing function and a completely
time-independent preferential-attachment-type growth kernel. For the general
case of such a separable citation rate, we obtain the time-dependent citation
distribution analytically in a form that is valid for any functional form of
its ageing and growth parts. Good agreement between theory and long-time
characteristics of patent-citation data establishes our work as a useful
framework for addressing still open questions about knowledge-propagation
dynamics, such as the observed excess of citations at short times.

\end{abstract}

\maketitle

\section{Introduction \& Overview of main results}

The structure and evolution of information transfer in collaborative
environments continues to be the subject of intense study. In particular,
citations by scientific articles~\cite{redner1998,borner2004,redner2005,
radicchi2008,golosovsky2012,golosovsky2013} and patents~\cite{Griliches1990,
jaffe1999,vonWartburg2005,csardi2007,valverde2007,sheridan2012} are being
investigated as directly accessible and suitably quantifiable indicators of
intellectual connectivity. Basic insight into citation data has been obtained
by the application of advanced models for network growth~\cite{sollaprice1976,
barabasi1999,dorogovt2000a,krapivsky2001,albert2002,newman2003} where the
principle of preferential attachment governs the creation of new connections
(i.e., citations). However, such growth dynamics is counterbalanced by the
typically increased tendency towards obsolescence for old knowledge. As a
result, the intrinsic~\footnote{The intrinsic citation rate $\lambda(t)$ is
obtained from the bare total citation rate $\lambda_\mathrm{tot}(t)$ by a
rescaling to account for the extrinsic variation in citability due to the
changing number $N(t)$ of patents that are generated at time $t$: $\lambda(t)
= \lambda_\mathrm{tot}(t)\, N(t_\mathrm{a})/N(t)$. $t_\mathrm{a} = 0.5$~years
in our analysis.} citation rate $\lambda(t)$ for patents (articles) at time $t$
can be surmised to be of the general form~\cite{csardi2007,valverde2007}
\begin{equation}\label{eq:sepRate}
\lambda(t) \equiv \bar\lambda(k(t), t) = A(t)\, f(k) \quad .
\end{equation} 
Here $A(t)$ is the ageing function~\cite{dorogovt2000b,zhu2003,Medo2011,
Wu2014}, and $f(k)$ embodies preferential-attachment dynamics through its
asymptotic power-law dependence $f(k) \sim k^\alpha$ on an individual patent's
(article's) cumulative number of citations $k(t)$.

\begin{figure*}
\includegraphics[width=0.33\textwidth]{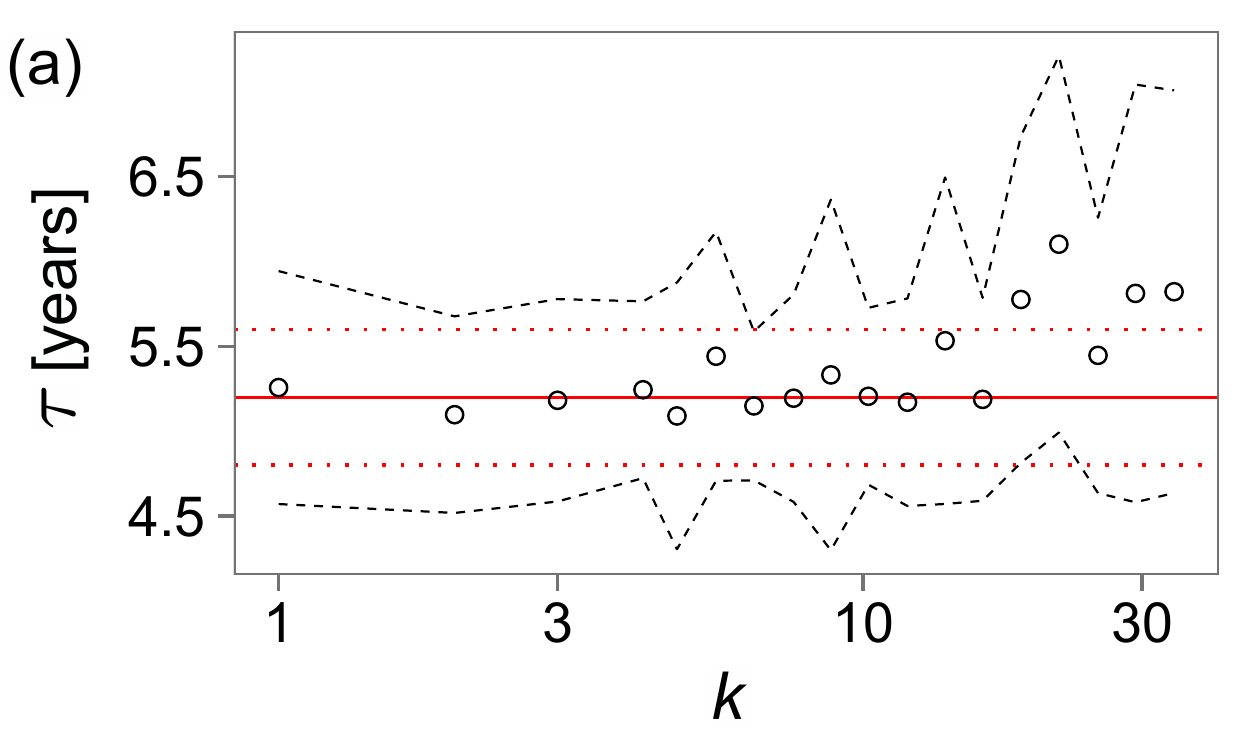}\hfill
\includegraphics[width=0.33\textwidth]{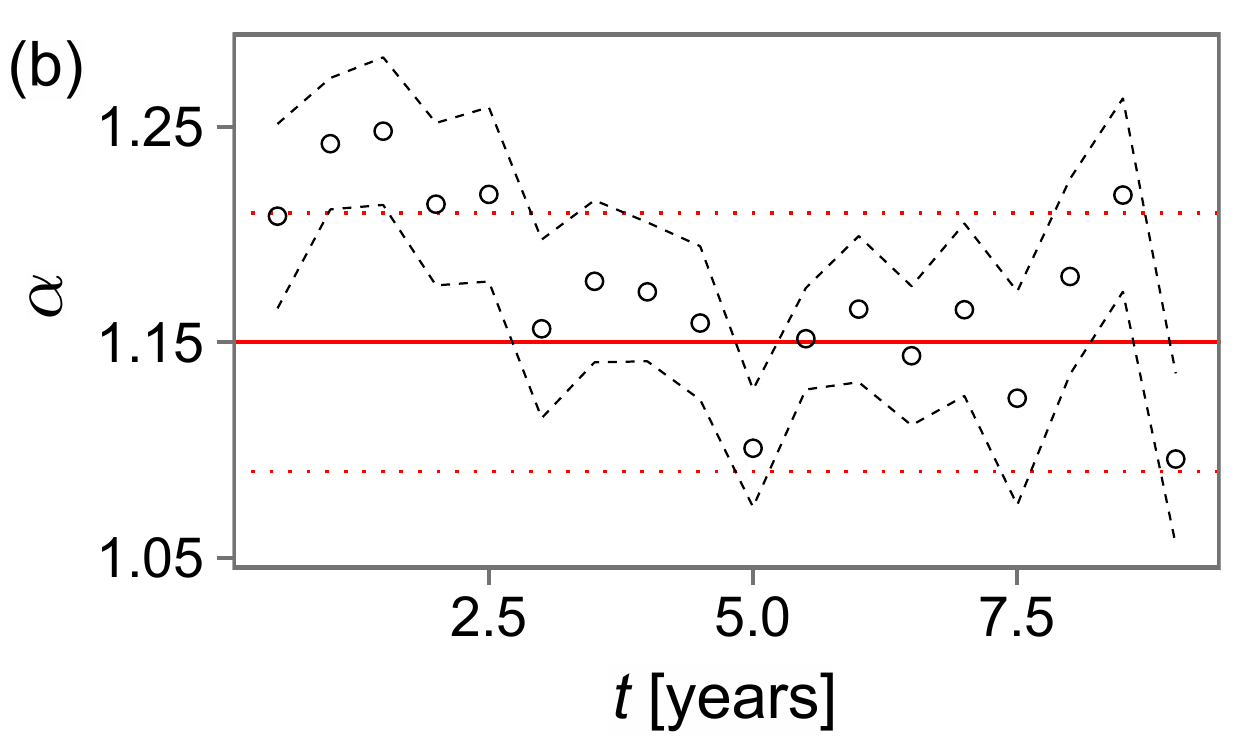}\hfill
\includegraphics[width=0.33\textwidth]{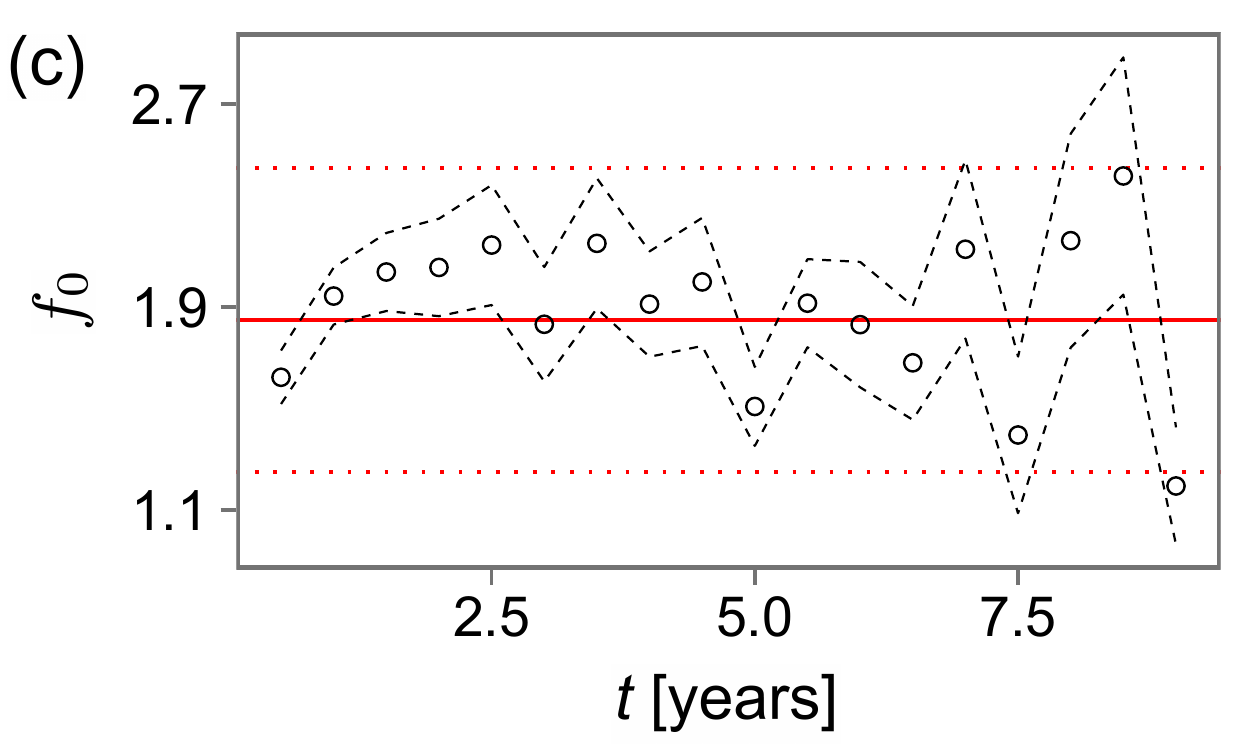}
\vspace{-0.4cm}
\caption{\label{fig:parameters}%
Disentangling preferential-attachment-type growth from ageing in the citation
rate of category-4 USPTO patents granted in 1998. (a)~Obsolescence time
$\tau$ extracted for each bin of patents with fixed number of citations $k$ from
fitting the citation rate to the form $\bar\lambda(k,t) \propto \exp(-t/\tau)$
for time $t\ge t_0 = 2.5$~years. (b)~Preferential-attachment exponent
$\alpha$, extracted for fixed times $t$ from fits of the citation rate to the
form $\bar\lambda(k,t) \propto (k^\alpha + f_0)$. (c)~Inhomogeneous
contribution $f_0$ to the preferential-attachment part of the citation rate,
extracted for fixed times $t$ from fits to $\bar\lambda(k,t) \propto (k^\alpha
+ f_0)$. Circles are fit-parameter values, solid lines their weighted averages,
and black dashed (red dotted) curves show 95\% confidence intervals for
fit-parameter values (weighted averages).}
\end{figure*}

\begin{table*}
\begin{tabular}{|c|cccccc|}
\hline
Category & 1 & 2 & 3 & 4 & 5 & 6 \\ \hline
$\alpha$ & $1.14 \pm 0.06$ & $1.13 \pm 0.06$ & $1.14 \pm 0.05$ & $1.15 \pm
0.06$ & $1.20 \pm 0.10$ & $1.20 \pm 0.07$ \\
$f_0$ & $1.3 \pm 0.3$ & $1.7 \pm 0.4$ & $1.0 \pm 0.2$ & $1.8 \pm 0.6$ & $2.0
\pm 0.9$ & $2.1 \pm 0.6$ \\
$\tau$ [years] & $5.6 \pm 0.8$ & $5.1 \pm 1.2$\footnote{\centering{Here $\tau$
showed a moderate residual dependence on the cumulative number of citations.}}
& $5.4 \pm 1.0^\mathrm{a}$ & $5.2\pm 0.4$ & $5.9 \pm 0.6$ & $5.9 \pm 0.5$ \\
$A_0$ [years$^{-1}$] & $0.29\pm 0.03$ & $0.28\pm 0.04$ & $0.40\pm 0.05$ &
$0.27\pm 0.03$ & $0.22\pm 0.02$ & $0.22\pm 0.02$ \\
$\Gamma_0$ &  0.92 & --~\footnote{\centering{Uncertainties in this category's
ageing-function parameters prevent determination of $\Gamma_0$.}} &
--$^\mathrm{b}$ & 0.87   &  0.94   &  0.91 \\ \hline
\end{tabular}
\caption{\label{parametertable}%
Parameter values for exponential ageing ($\tau$, $A_0$) and
preferential-attachment-type growth ($\alpha$, $f_0$), extracted for patent
cohorts from individual technology categories according to the classification
of Ref.~\cite{hall2003} where 1 = Chemical, 2 = Computers \& Communications,
3 = Drugs \& Medical, 4 = Electrical \& Electronic, 5 = Mechanical, and 6 =
Others. The parameter $\Gamma_0$ quantifies the departure from exponential
ageing at short times through its deviation from $1$.}
\end{table*}

Various specific functional forms for $A(t)$ and $f(k)$ have been fitted to
real citation data. For the \textit{Ansatz\/} (\ref{eq:sepRate}) to be
meaningful, fitting procedures should find that any parameters entering the
preferential-attachment kernel $f(k)$ are independent of time once an accurate
model for ageing of knowledge has been adopted and extrinsic influences, such as
the increased number of patents (articles) generated over time, have been
controlled for by a suitable normalisation. Here we show that this is indeed the
case for patents within given technology categories. More specifically, the
validity of (\ref{eq:sepRate}) is demonstrated by explicit extraction of the
ageing function and preferential-attachment kernel, which are found to be of the
form
\begin{subequations}
\begin{eqnarray}\label{eq:ageing}
A(t) &=& A_0\, \exp\left(-\frac{t}{\tau}\right) \quad \mbox {for $t \ge t_0
\approx \tau/2$} \,\, , \\[0.1cm] \label{eq:prefatt}
f(k) &=& k^\alpha + f_0 \,\, .
\end{eqnarray}
\end{subequations}
Figure~\ref{fig:parameters} shows results of a fitting procedure that most
directly demonstrates the validity of the separation \textit{Ansatz\/}
(\ref{eq:sepRate}), and the parameter values found for individual technology
categories defined in Ref.~\cite{hall2003} are given in
Table~\ref{parametertable}.

Our present approach~\footnote{Three aspects -- disaggregation by technology,
citation-inflation adjustment, and use of the time lag between granting of a
cited patent and the \emph{application date} of citing patents as the relevant
time parameter -- differentiate our present approach from the one followed by
previous studies of citation data~\cite{csardi2007,valverde2007,
golosovsky2012}.} establishes the rate expression (\ref{eq:sepRate}) as a
viable description for patent-citation dynamics and so provides the means to
investigate general aspects of knowledge flow in greater detail. As an example,
we discuss interesting insights that emerge from comparing values for
growth-kernel and ageing-function parameters associated with the citation
statistics for patents from different technology categories. On a conceptual
level, finding ageing to be reliably modeled by an exponential function of
time in the long run implies the existence of a finite life time $\tau$ for
patents before they become obsolete -- in agreement with an early study of
patent-citation dynamics~\cite{jaffe1999} that also observed exponential
ageing~\footnote{More recent works~\cite{csardi2007,golosovsky2012} postulated
ageing functions with asymptotic power-law behavior $A(t)\sim t^{-\beta}$ but
concomitantly observed a marked increase of the preferential-attachment
exponent $\alpha$ over time. Others \cite{valverde2007} presumed the ageing
function to be of Weibull form but fixed $\alpha=1$ in their fits. As shown in
Appendix~\ref{app:alternatives}, exponential ageing best describes the intrinsic
citation rate for patents in the long run.}. Systematic deviations from the, in
the long-term very accurate, exponential-aging model point to currently not
understood mechanisms for knowledge propagation at short times that require
further study. 

As one of our main results, we obtain the fully general analytic expression for
the distribution function $n(k, t)$ for citations, which is the fraction of
patents having $k$ citations at time $t$. For all values $k\ge 1$, it will
satisfy the Master
equation~\cite{Medo2011}
\begin{subequations}
\begin{equation}\label{eq:finiteCit}
\frac{d n(k, t)}{dt} = -\bar\lambda(k, t)\, n(k, t) + \bar\lambda(k-1, t)\,
n(k-1, t)\,\, ,
\end{equation}
whereas for $k=0$, the corresponding Master equation is
\begin{equation}\label{eq:zeroCit}
\frac{d n(0, t)}{dt} = -\bar\lambda(0, t)\, n(0, t) \quad .
\end{equation}
\end{subequations}
Solving these equations for the most general form of initial conditions $n(k,0)
= n_0(k)$ allows us to account for the fact that a significant number of patents
have already acquired citations by the time they are granted. Assuming only
that the citation rate has the form given in (\ref{eq:sepRate}), we 
find~\footnote{Solution of Eq.~(\ref{eq:zeroCit}) yields $n(0,t) = n_0(0)
\left[ \gamma(t) \right]^{f(0)}$. For given $n(k-1, t)$,
Eq.~(\ref{eq:finiteCit}) is solved by the method of variation of constants,
yielding $n(k, t) = n_0(k) \left[ \gamma(t) \right]^{f(k)}+\left[ \gamma(t)
\right]^{f(k)} \int_0^t dt' \,\, \bar\lambda(k-1, t')\, \left[\gamma(t')
\right]^{-f(k)} \, n(k-1, t')$. The explicit result given in
Eq.~(\ref{eq:citDistrib}) is obtained by induction.}
\begin{subequations}
\begin{eqnarray}\label{eq:citDistrib}
n(k, t) &=& n_0(k) \left[\gamma(t)\right]^{f(k)} \nonumber \\[0.2cm]
&& \hspace{-1.3cm} + \sum_{l=0}^{k-1} n_0(l) \left( \prod_{m=l}^{k-1} f(m)
\right) \sum_{q=l}^k \frac{\left[\gamma(t)\right]^{f(q)}}{\prod_{m=l \atop
m\ne q}^k \left[f(m) - f(q) \right]} , \quad 
\end{eqnarray}
where
\begin{equation}\label{eq:gamma}
\gamma(t) = \exp\left\{-\int_0^{t} dt'\, A(t')\right\} \quad .
\end{equation}
\end{subequations}
This result is applicable to any citation dynamics described by a rate that is
the product of an ageing part and a citation-number-dependent growth part,
irrespective of their explicit functional forms. It also incorporates the fully
general initial conditions.

\begin{figure}[b]
\includegraphics[width=0.7\columnwidth]{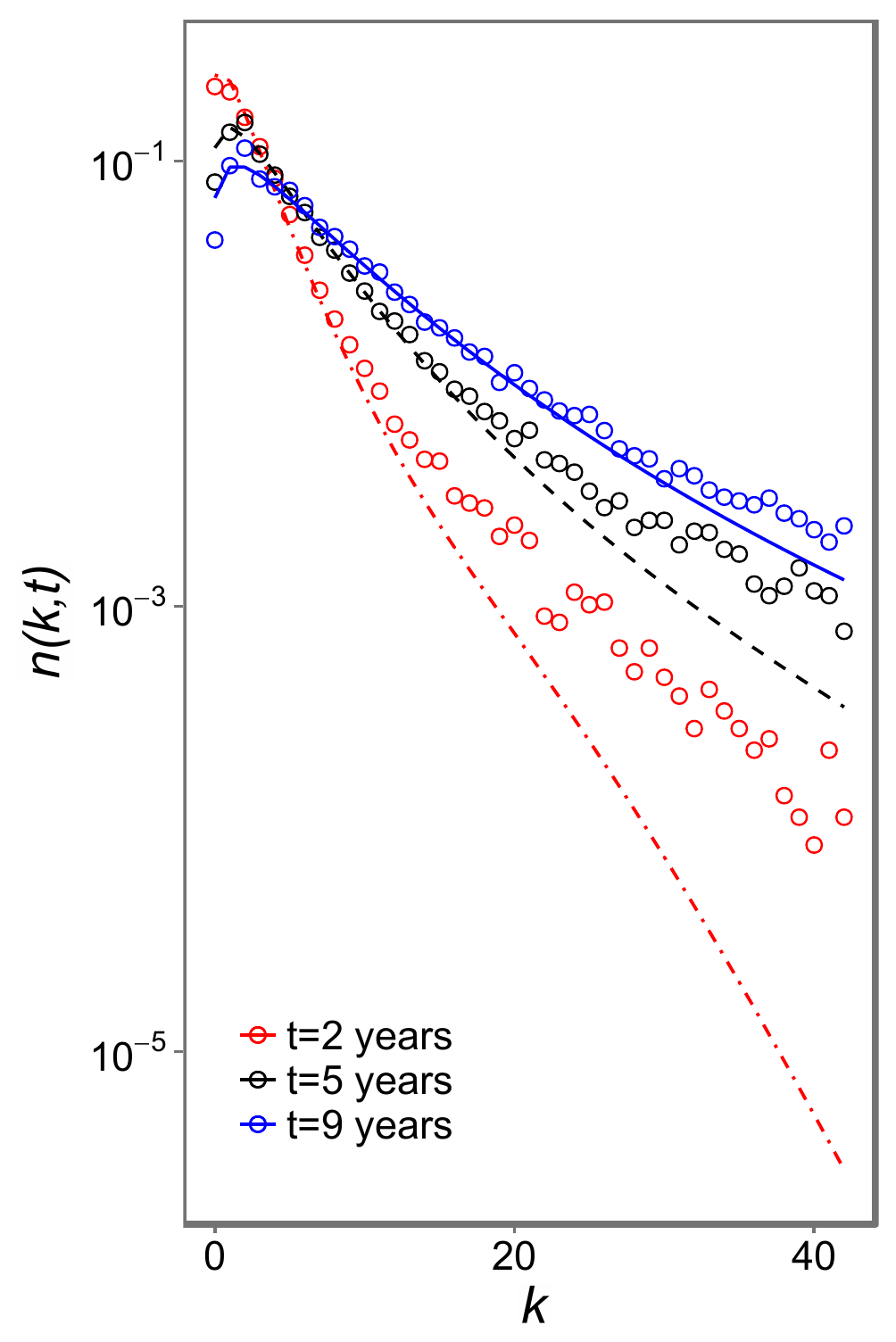}
\vspace{-0.4cm}
\caption{\label{fig:distriComp}%
Distribution function $n(k, t)$ for patent citations, plotted as a function of
the number $k$ of citations for fixed times $t = 2$, 5 and 9 years. Empirical
data for the cohort of category-4 patents granted in 1998 are shown as symbols.
Solid curves have been calculated from the general theoretical expression in
Eq.~(\ref{eq:citDistrib}) with functional forms for the ageing function and the
growth kernel given in Eqs.~(\ref{eq:ageing}) and (\ref{eq:prefatt}),
respectively. To capture deviations from exponential ageing at short times, the
factor $\Gamma_0$ has been introduced as explained in the text. Parameters used
are those listed for category 4 in Table~\ref{parametertable}.}
\end{figure}

Specializing (\ref{eq:gamma}) to the ageing function given in (\ref{eq:ageing})
yields
\begin{subequations}
\begin{equation}
\gamma(t) = \Gamma(t) \exp\left\{ \tau A_0 \left[ \exp \left( -\frac{t}{\tau}
\right) - 1 \right] \right\} \quad ,
\end{equation}
where
\begin{equation}
\Gamma(t)\equiv \exp \left\{ - \int_0^{t} dt'\, \left[ A(t') - A_0 \exp \left(
-\frac{t'}{\tau} \right)\right]\right\}
\end{equation}
\end{subequations}
quantifies the effect of deviations from exponential ageing that occur at short
times $t\le t_0$. Hence we expect, and indeed observe, $\Gamma(t > t_0) \to
\Gamma_0 = \mathrm{const}$. For the purpose of the present work, extracting the
parameter $\Gamma_0$ from the data enables us to model correctly the
citation-number distribution $n(k,t)$ for $t\ge t_0$. See
Fig.~\ref{fig:distriComp} for a comparison of the theoretical prediction with
the data. Table~\ref{parametertable} summarizes the values of $\Gamma_0$ found
for individual technology categories.

\begin{figure}[b]
\includegraphics[width=0.9\columnwidth]{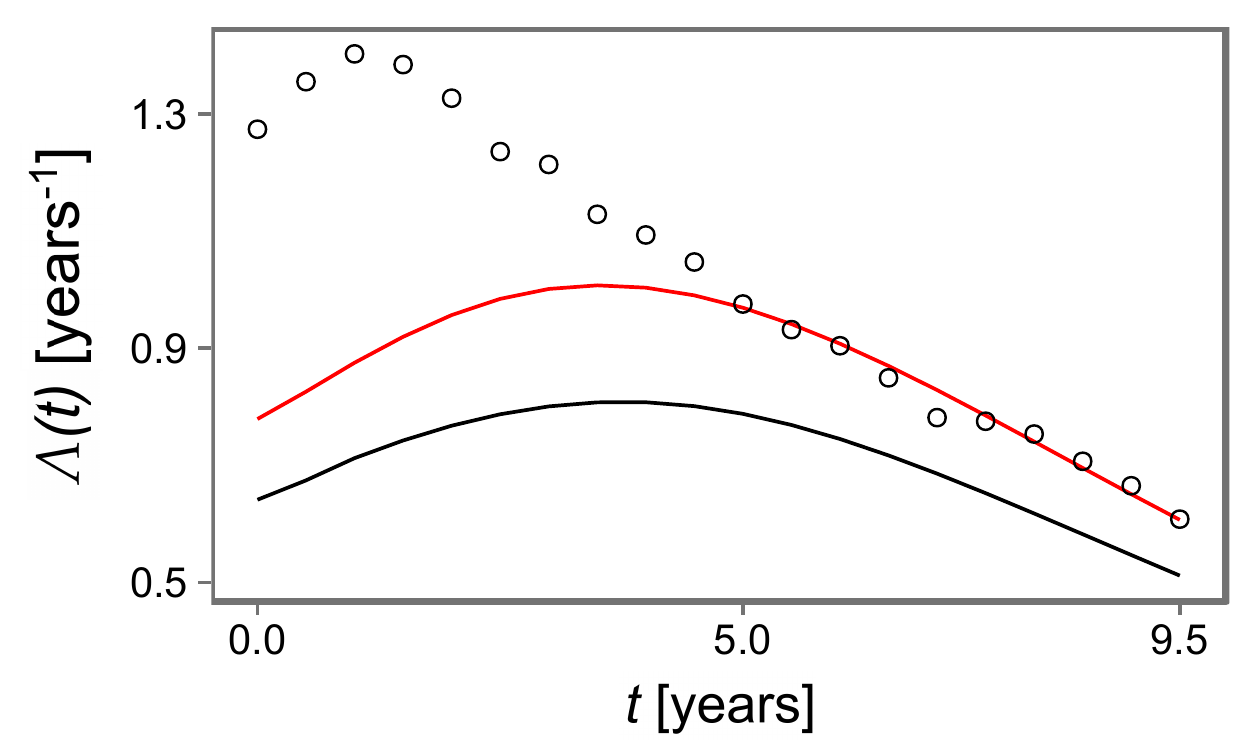}
\vspace{-0.4cm}
\caption{\label{fig:avrgRate}%
Average citation rate $\Lambda(t)$ for category-4 patents granted by USPTO in
1998. Symbols represent the data, and the red (black) curve is the theoretical
prediction based on applying preferential-attachment-type growth and
exponential ageing to the measured initial citation distribution $n_0(k)$, with
(without) correcting for short-time deviations from exponential ageing using
$\Gamma_0 = 0.87$.}
\end{figure}

Knowing $n(k,t)$ explicitly, it is possible to calculate any citation-related
quantity of interest, including the average rate of citation for patents from
the cohort; $\Lambda(t) =\sum_k n(k, t)\, \bar\lambda(k, t)$, and the average
cumulative number of citations obtained over time; $K(t)=\sum_k n(k, t)\,
k$~\footnote{For $\alpha=1$, $f_0=1$, $n_0(k)=\delta_{k0}$ as assumed, e.g., in
Refs.~\cite{Medo2011,Wu2014}, the result (\ref{eq:citDistrib}) specializes to
$n(k,t)=\gamma(t) \left[ 1 - \gamma(t)\right]^k$, yielding $\Lambda(t)=A(t)/
\gamma(t)$ and $K(t)= \left[ \gamma(t)\right]^{-1} - 1$.}, which are related
via $d K(t)/dt =\Lambda(t)$. We only consider $\Lambda(t)$ here.
Figure~\ref{fig:avrgRate} shows a representative comparison between the
average citation rate obtained from actual patent-citation data and the
prediction coming out of our model that combines preferential-attachment-type
growth with ageing. For the theoretical curves, we have taken into account the
empirically measured initial citation distribution $n_0(k)$. The assumption of
purely exponential ageing turns out not to yield a good description of the data
for $\Lambda(t)$, but good agreement at longer times is achieved simply by
setting $\Gamma(t)\equiv \Gamma_0$ with the appropriate value for $\Gamma_0$ to
account for the effect of short-time deviations from exponential ageing. Thus
despite the fact that the exact functional form of the ageing function $A(t)$
at short times $t< t_0$ is currently not known, we can accurately model the
longer-term behavior of $\Lambda(t)$.

\begin{table*}
\begin{tabular}{|c|cccccc|}
\hline
Technology category (after \cite{hall2003}) & 1 & 2 & 3 & 4 & 5 & 6 \\ \hline
Number of patents granted in 1998 & 32,626 & 34,872 & 20,703 & 35,527 & 32,423 &
33,275 \\
Total citations accrued by cohort & 333,306 & 809,396 & 406,373 & 549,216 &
360,609 & 349,081 \\ 
Total of inflation-adjusted citations & 301,451 & 606,489 & 345,458 & 435,430 &
316,326 & 322,515 \\ \hline
\end{tabular}
\caption{\label{statstable}%
Summary statistics for patent dataset. We consider cohorts formed by patents
granted in 1998 that belong to a given technology category and count citations
to these made by patents from any category with application date before 2008.
Besdies the raw total number of citations to each such cohort, we also give the
total obtained by summing the normalized values from each year that were
adjusted to account for citation inflation as described in the text.}
\end{table*}

The remainder of this Article is organized as follows. We describe the data
and methods used for our analysis in Sec.~\ref{sec:methods}. General
implications of our results for understanding patent-citation dynamics and
associated knowledge characteristics are discussed in
Sec.~\ref{sec:discussion}. We present conclusions in Sec.~\ref{sec:concl} and
provide relevant additional information in Appendices.

\section{Methodology for data analysis \& fitting}\label{sec:methods}

Our data comprise the citation history of all patents granted in the year 1998
by the United States Patent and Trademark Office (USPTO), where each patent is
assigned one of six broad technology categories as laid out in
Ref.~\cite{hall2003}. The cohort of patents granted in 1998 was chosen in order
to have both enough data and enough time to conduct a robust analysis of the
network evolution. The date of a citation is taken to be the application date
of the citing patent. We consider all citations to this cohort up to, and
including, those applied for in the year 2007. A detailed summary statistics
for our patent-citation dataset is provided in Table~\ref{statstable}.

The citations to each patent in our cohort are binned in six-month periods,
starting from the grant date of the patent. This establishes two time series;
one for the number of additional citations gained within a given six-month
period, and one for the total number of citations that have been accrued by the
end of each period. These citations are adjusted for `citation inflation' to
mitigate extrinsic factors affecting the citation rate: because the number of
patent applications in each period and technology category is variable, we
divide the nominal application counts by the number of applications in 1998 to
scale each citation to an `equivalent 1998 citation'. We perform this scaling
only within a given technology category based on the approximation that patents
mostly cite other patents from the same technology category. The fact that we
see consistent exponential aging behavior across most technology categories
suggests that this approximation is generally valid. See Table~\ref{statstable}
for a listing of both the raw total and `inflation-adjusted' citation numbers
for each cohort of patents.

We also bin the data into groups of patents with similar values of $k$ at $t$.
In order to ensure equal spacing of these bins during the modeling process, we
implement logarithmic binning, whereby bin size gets exponentially larger with
the number of cumulative citations. There are few patents with a large number
of citations, and so the measured citations for these patents have large
uncertainties. To mitigate this, we define a threshold, and when the number of
accrued citations to a patent exceeds this threshold, the patent gets excluded
from the analysis. For our present study, the threshold has been set at the
95th percentile of total accrued citations, which varies with technology
category.

\begin{figure}[t]
\includegraphics[width=0.9\columnwidth]{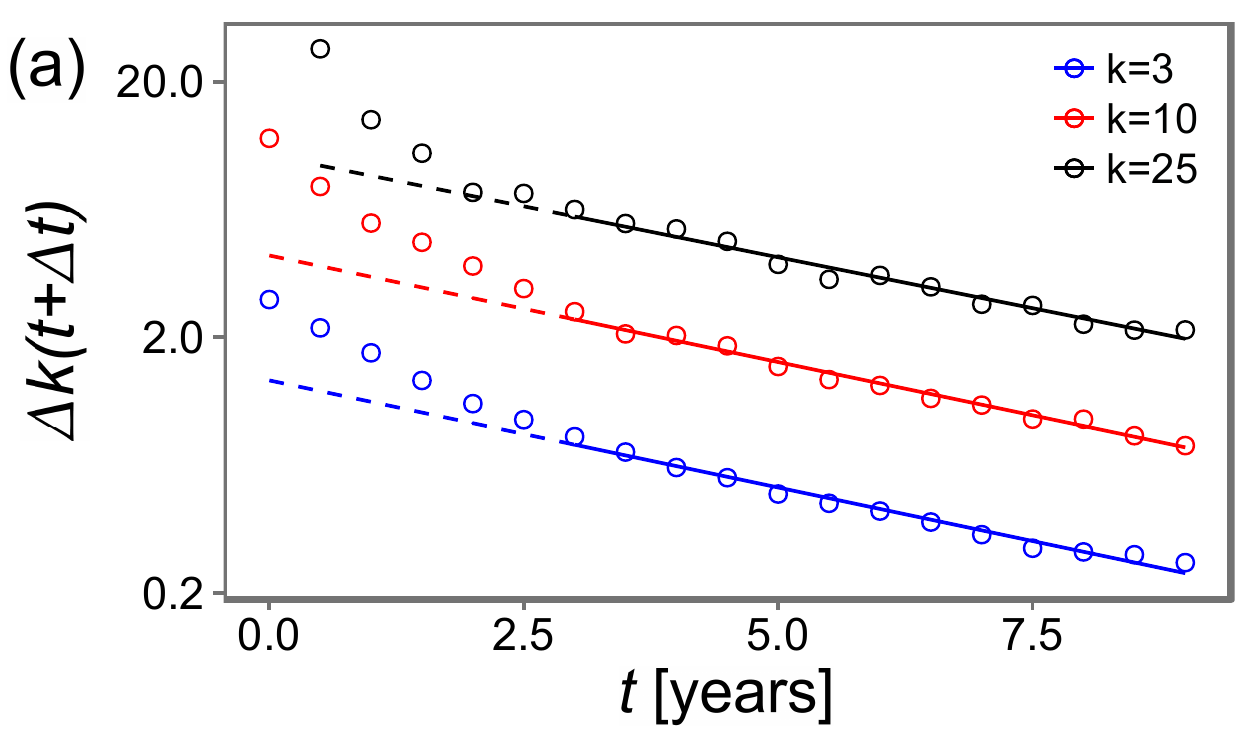}\\
\includegraphics[width=0.9\columnwidth]{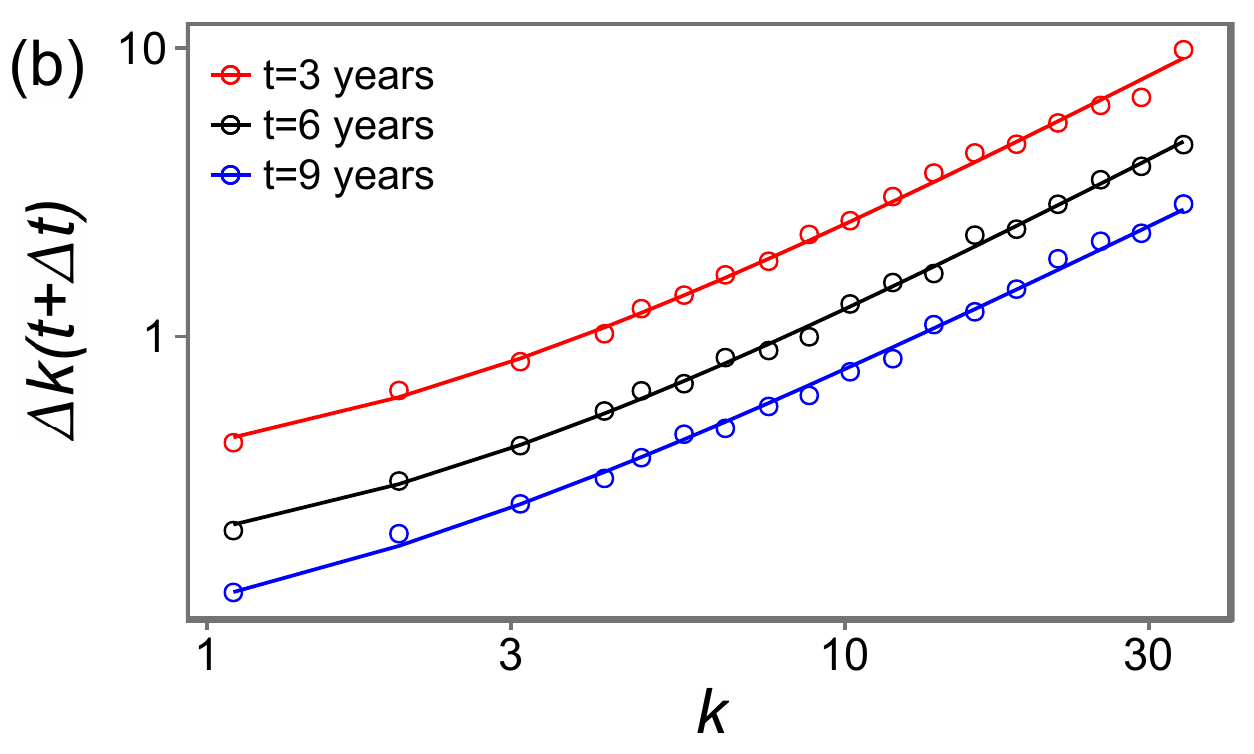}\\
\caption{\label{fig:fits}%
Fitting the time and citation-number dependences of the empirical citation rate.
(a)~Mean number of additional citations, $\Delta k(t+\Delta t)$, as a
function of time $t$ for fixed citation bins with log-midpoints at $k=3$, 10
and 25. Hollow circles are derived from the citation data, while solid lines
represent the fit to the form of  Eq.~(\ref{eq:sepRate}) with $A(t)$ given by
Eq.~(\ref{eq:ageing}). Deviations from exponential-ageing behavior occur only
at short times. (b)~Mean number of additional citations, $\Delta k
(t+\Delta t)$, as a function of cumulative number of citations $k(t)$ for
various fixed times $t=3$, 6 and 9 years. Hollow circles are derived from the
citation data, while solid lines represent the fit to the form of
Eq.~(\ref{eq:sepRate}) with $f(k)$ given by Eq.~(\ref{eq:prefatt}).}
\end{figure}

To prove separability of the citation rate and extract relevant parameters, we
assume, for patents with $k$ citations, that the average additional number of
citations in $(t,t+\Delta t]$, denoted $\Delta k(t+\Delta t)$, with $\Delta t =
6$~months is a good proxy for the citation rate at time $t$. (As shown in
Appendix~\ref{app:robust_dt}, our results do not depend on this particular
convenient choice of $\Delta t$.) We then analyze how these rates change as
functions of both time and accumulated citations. In particular, we fit
Eq.~(\ref{eq:ageing}) to the time dependence of the observed citation rate at
various fixed $k$ to extract values of $\tau$ [see Fig.~\ref{fig:fits}(a)], and fit
Eq.~(\ref{eq:prefatt}) to the $k$ dependence of the observed citation rate at
various fixed $t$ to extract values of $\alpha$ and $f_0$ [cf.\ Fig.~\ref{fig:fits}(b)].
All fits performed in this
work are carried out using a weighted nonlinear least-squares regression with
weights proportional to the number of patents represented by each data point.
We fit (\ref{eq:ageing}) to times $\gtrsim\tau/2$ because of a departure from
exponential behaviour observed at short times. The results obtained in that way
for patents from category 4 (Electrical \& Electronic) are shown in
Fig.~\ref{fig:parameters}. The fact that there is no systematic residual time
dependence in the values for $\alpha$ and $f_0$, and no dependence of the
extracted $\tau$ on the number of citations $k$, clearly shows the
applicability of the separation \textit{Ansatz} (\ref{eq:sepRate}) for the
citation rate. Repeating the fitting procedure using potential alternative
forms for $f(k)$ and $A(t)$, as discussed in Appendix~\ref{app:alternatives},
establishes Eqs.~(\ref{eq:ageing}) and (\ref{eq:prefatt}) as the most
meaningful and accurate models. This allows us to discuss more systematic
specifics of the ageing and growth mechanisms.

\section{Discussion of results \& Outlook}\label{sec:discussion}

\subsection{Distribution function for patent citations:\\ Comparison between
theory and data}

The distribution function $n(k, t)$ captures all aspects of patent-citation
statistics and is therefore an extremely useful quantity to know. We have
obtained its fully general analytic expression in Eq.~(\ref{eq:citDistrib})
for any situation where the citation rate has a separable form as given in
Eq.~(\ref{eq:sepRate}). Results obtained from calculating this expression using
the functional forms $f(k) = k^\alpha + f_0$ and $$\gamma(t) = \Gamma_0 \exp
\left\{ \tau A_0 \left[ \exp \left( -\frac{t}{\tau} \right) - 1 \right] 
\right\}$$ with the parameters we have extracted from analyzing citation data
for category-4 patents granted in 1998 are plotted in
Fig.~\ref{fig:distriComp}. Comparison with the empirical data for $n(k, t)$
reveals the regions of validity for our model. At large-enough $t$, the
agreement is very good over essentially the entire range of $k$. As $t$ gets
smaller, the region of good agreement shrinks to a limited range of smaller and
smaller $k$. Based on this understanding of the properties of the distribution
function, the accuracy of our model for calculating any statistical quantity of
interest related to patent citations can be inferred.

\subsection{Robust preferential-attachment growth}

Within our analysis, we observe no systematic deviations from
preferential-attachment-type growth. Our extracted values of the exponent
$\alpha$ vary moderately between the different technology categories but are all
broadly consistent with $\alpha\gtrsim 1.15$. This value is marginally lower
than those typically found previously~\cite{golosovsky2013} from citation
networks, but still implies super-linear growth dynamics. The initial
attractiveness for a patent to be (essentially randomly) cited is quantified by
the parameter $f_0$ whose values are in the sensible range $1\lesssim f_0 
\lesssim 2$ across all technology categories. The observed
technology-independent universality of the preferential-attachment growth
behavior points to the existence of common drivers for citation-based knowledge
flow in invention space.

\subsection{Technology dependence of short \& long-term ageing}

Our study has succeeded in establishing the separability of the citation rate
into ageing and growth parts, and has also clarified the exact form of the
preferential-attachment kernel. However, the functional form of the ageing
function $A(t)$ was able to be reliably identified as being exponential only for
longer times $t\ge t_0 \approx 2.5$~years. As illustrated in Fig.~\ref{fig:fits}(a),
the citation rate at shorter times is larger than expected
from purely exponential ageing with the extracted lifetime $\tau$. Nevertheless,
the knowledge of exponential ageing occurring in the long run, together with the
separability according to (\ref{eq:sepRate}), make it possible to at least find
an accurate model for the distribution $n(k,t)$ of citations for times $t\ge
t_0$ (see Fig.~\ref{fig:distriComp}), which can be the starting point for
further study into the short-time ageing mechanism.

Similarly to the preferential-attachment parameters, the life time associated
with long-term exponential ageing turns out to show some variation across the
different technology categories. The broad range $5\lesssim \tau/\mathrm{years}
\lesssim 6$ is consistent with early results reported in the economics
literature~\cite{jaffe1999}. Furthermore, those technologies that are generally
perceived to be faster-changing (categories 2-4) exhibit shorter $\tau$ while
those perceived to change more slowly (categories 1, 5 and 6) exhibit longer
$\tau$. See Table~\ref{parametertable}. The factor $\Gamma_0$ that
quantitatively embodies short-term deviations from exponential ageing also
exhibits an interesting pattern across technologies. It is plausible that the
enhancement of knowledge transfer that is observed for some technologies on
short time scales does not occur to a significant level for patents from the
very heterogeneous and generally slower-changing technology space of categories
1, 5 and 6. This leads to a larger $\Gamma_0 \gtrsim 0.9$ for these categories.
In contrast, the faster-developing category 4 is strongly influenced by such
short-time effects, rendering its $\Gamma_0$ to be smaller.

\subsection{Relating patent-citation statistics to knowledge flow: Opportunities
and caveats}

We treat citations as an indicator or proxy for knowledge flow. We might expect
citations in patents to be a high-quality indicator, because they are governed
by explicit legal rules, and their inclusion in or exclusion from a given
patent has legal consequences~\cite{jaffe2016}. Indeed, recent work uses
transitive reduction (TR) ~\footnote{TR removes links without disrupting
`information flow', such that, when patent A cites patents B and C, and B also
cites C, then the edge connecting A to C is removed, as the information still
flows from C to A via B. This edge-removal process is argued to highlight the
causal structure of a directed acyclic network, for which the resulting graph
is unique. The authors of Ref.~\cite{clough2015} find that TR removes about
80\% of links in academic citation networks, but only 15\% in patent-citation
networks.} of citation networks to show that the level of redundancy in the
information flow associated with patent citations is much smaller when compared
with citations to academic papers ~\cite{clough2015}. This suggests a stronger
link between citation dynamics and knowledge flow in the patent network as
compared to networks mapped from article citations (for which both rules and
consequences are much less clear). There are, however, reasons to be cautious
in interpreting patent citations as indicators of knowledge flow. Citations are
made by several parties (including the inventor, the patent attorney, and the
USPTO patent examiner~\footnote{This is illustrated by a large survey of
inventors~\cite{jaffe2000} finding that almost half of the citations on their
patents referenced inventions unknown to the inventors before the survey.}) and
for various reasons~\cite{cotropia2013}. Applying our present approach to
analyze the statistics of patent citations that originate from different
parties acting during the patenting process could shed interesting new light on
these various influences~\cite{jaffe2016}.

\section{Conclusions}\label{sec:concl}

We have empirically established that the intrinsic rate of citation for patents
within individual technology categories is separable into an ageing function and
a preferential-attachment growth kernel. The ageing function depends
exponentially on time in the long run, but the functional form characterizing
its short-time behavior is currently not known. The life time associated with
the exponential ageing and the parameters of preferential-attachment-type growth
have been reliably extracted, and variations across technology categories were
discussed. Based on the separability of the intrinsic citation rate, the
distribution of citations has been obtained as a function of time in an
explicitly analytic form that also incorporates a completely general initial
citation distribution. This result enables detailed modelling of patent-citation
statistics and facilitates future in-depth investigations into mechanisms for
knowledge propagation through invention space.

\appendix

\section{Robustness of data analysis with respect to a change in sampling period
$\mathbf{\Delta t}$ for incoming citations}\label{app:robust_dt}

For the analysis presented in the main text, we have counted incoming citations
received over the interval $(t, t+\Delta t]$ with $\Delta t =6$~months to
determine the citation rate at time $t$. As Table~\ref{robustnesstable} shows,
different reasonable choices for the sampling period $\Delta t$ yield the same
parameters resulting from fitting.

\begin{table}[b]
\begin{tabular}{|c|ccc|}
\hline $\Delta t$ [months] & 6 & 4 & 3  \\ \hline
$\alpha$ & $1.15 \pm 0.06$ & $1.14 \pm 0.07$ & $1.16 \pm 0.08$ \\
$f_0$ & $1.8 \pm 0.6$ & $1.7 \pm 0.5$ & $1.7 \pm 0.9$ \\
$\tau$ [years] & $5.2\pm 0.4$ & $5.2 \pm 0.5$ & $5.2 \pm 0.4$ \\
$A_0$ [years$^{-1}$] & $0.27\pm 0.03$ & $0.28\pm 0.05$ & $0.27\pm 0.05$ \\\hline
\end{tabular}
\caption{\label{robustnesstable}%
Parameters obtained for category-4 patents granted in 1998 when incoming
citations are counted over periods of length $\Delta t = 6$, 4, and 3 months,
respectively. There is very good agreement between average values to within
uncertainties. As expected, parameter uncertainties increase as $\Delta t$ is
reduced because lower citation numbers over shorter periods result in larger
statistical errors.}
\end{table}

Generally, the optimal choice of $\Delta t$ will take account of the amount of
available data (cohort size; number of incoming citations) and the length of
time over which citation statistics is to be modeled. Choosing a larger $\Delta 
t$ is advisable, e.g., for analyzing cohorts with few patents and/or incoming
citations. In contrast, a shorter $\Delta t$ may be necessary when working with
more-recently-granted patents for which there has been less time to accrue
citations.

\begin{figure*}
\includegraphics[width=0.33\textwidth]{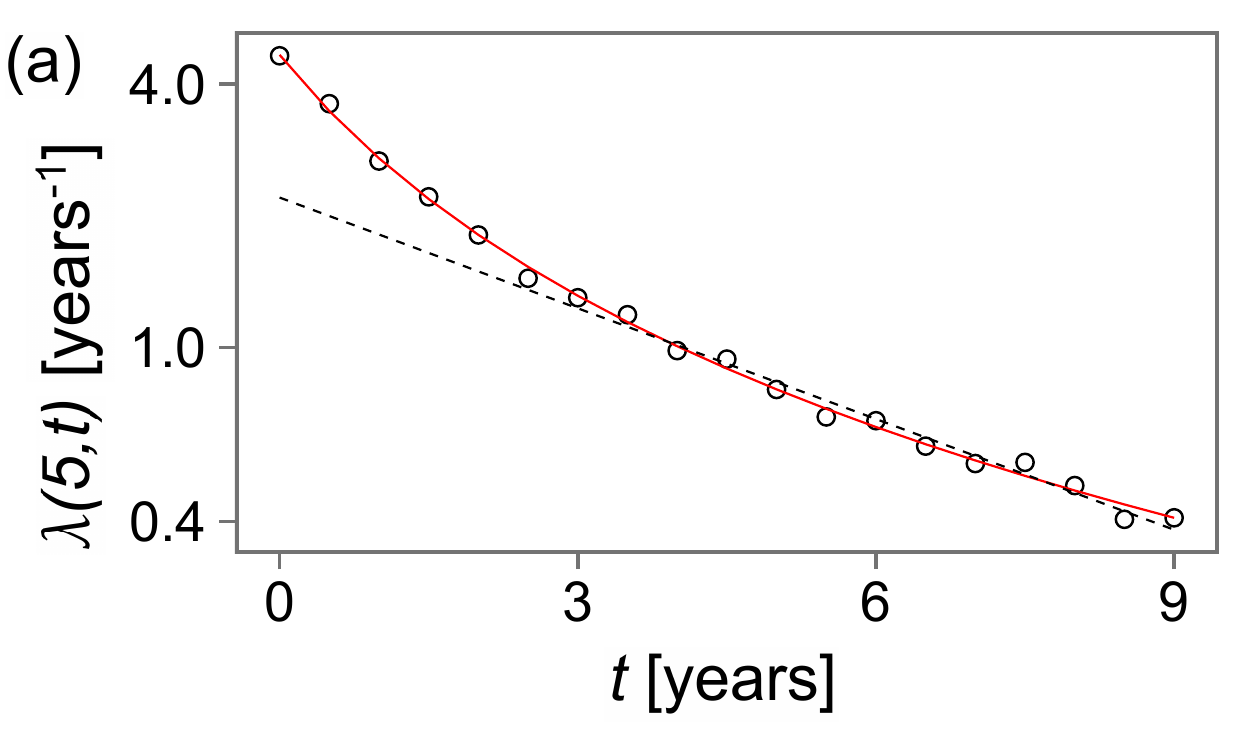}\hfill
\includegraphics[width=0.33\textwidth]{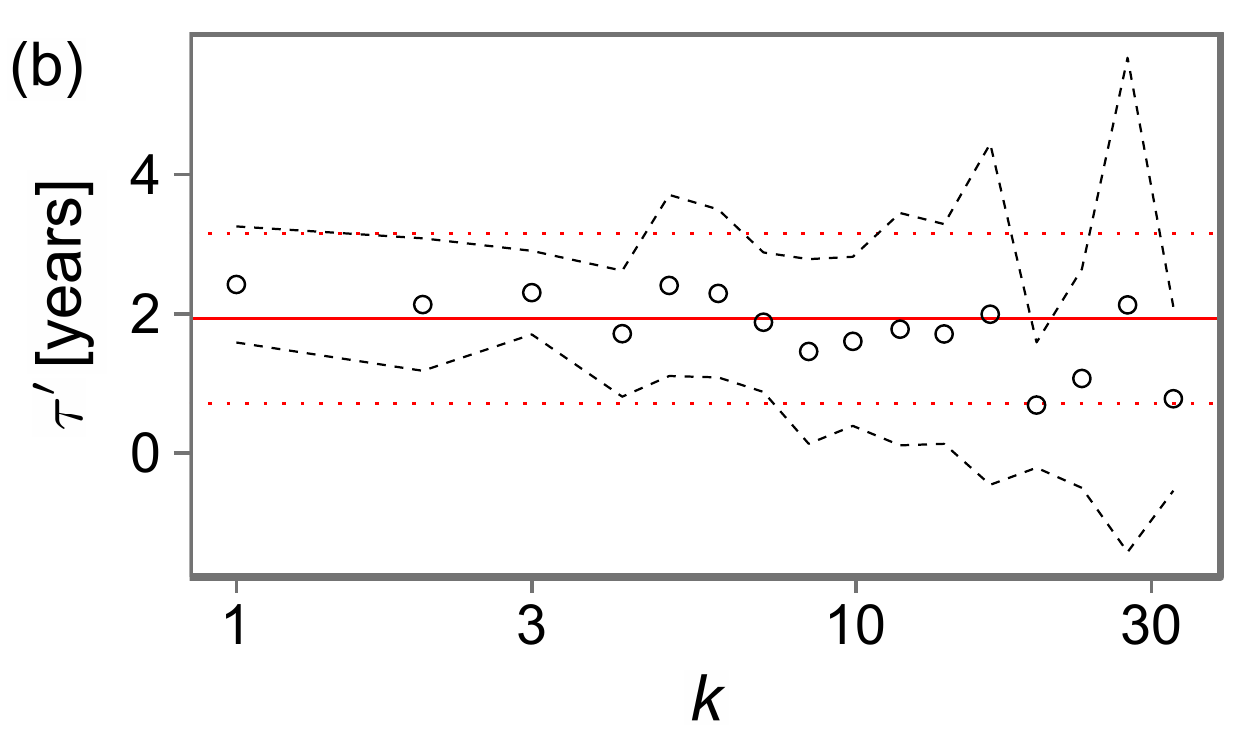}\hfill
\includegraphics[width=0.33\textwidth]{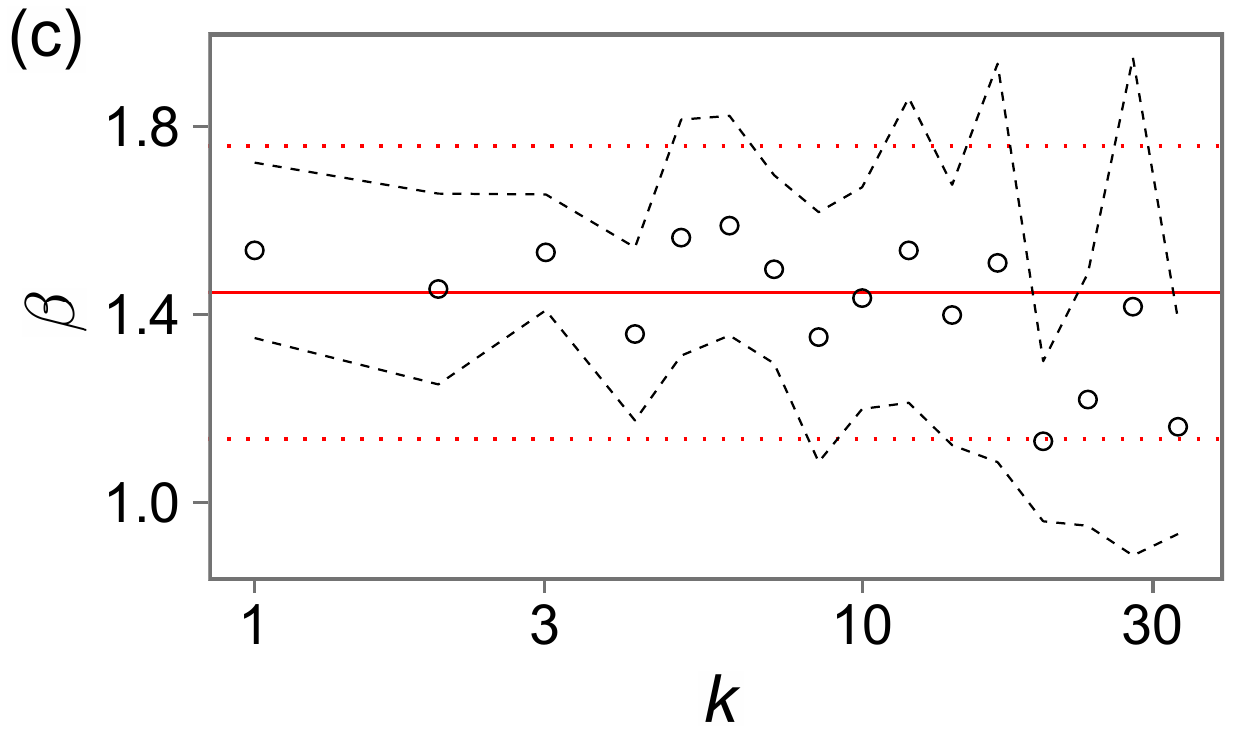}
\vspace{-0.4cm}
\caption{\label{fig:powerlaw}%
Power-law vs.\ exponential ageing: The three-parameter power law
(\ref{eq:powerAge}) fits data superficially well but yields very uncertain
parameter values. For the case $k=5$ shown in panel~(a), the power-law
fit (red solid curve) yields $\beta=1.57 \pm 0.25$ and $\tau' = (2.40 \pm
1.30)$~years; relative uncertainties of 16\% and 54\% respectively at a 95\%
confidence level. This can be contrasted with the exponential fit (black dashed
line) that gives $\tau = (5.14 \pm 0.61)$~years -- a relative uncertainty of
12\% -- while providing a fit that is basically indistinguishable from the
power-law fit for times $t\gtrsim\tau/2$. Parameters $\tau'$ and $\beta$
obtained from fitting the power-law ageing model at different fixed $k$ are
shown in panels (b) and (c), respectively. Note the consistently large
uncertainties and systematic residual $k$ dependence, which severely
limit the utility of the power-law ageing model (\ref{eq:powerAge}) for
analyzing our patent-citation data set. Solid red lines represents weighted
averages, and black dashed (red dotted) lines represent 95\% confidence
intervals for the fit-parameter values (weighted averages).}
\end{figure*}

\section{Exploring alternative functional forms describing growth and
ageing dynamics in the citation rate}\label{app:alternatives}

\subsection{Alternative preferential-attachment kernels}

While preferential-attachment models generally have a kernel that
asymptotically approaches a power law~\cite{krapivsky2001}, $\lim_{k\to\infty}
f(k) = k^\alpha$, different specific functional forms have been proposed to fit
real data. Here we provide a comparison between our adopted form $f(k) =
k^\alpha + f_0$, which was also used previously for patent
data~\cite{csardi2007}, and a possible alternative for $\alpha\ne 1$, $f(k) =
(k+k_0)^\alpha$, which has been applied to model citation growth for scientific
publications~\cite{golosovsky2012}. Table~\ref{preftable} shows results from
fitting the relevant parameters to the patent data. Values obtained for the
exponent $\alpha$ are quite close but have significantly lower uncertainties for
our adopted functional form, motivating our particular choice.

\begin{table}[b]
\begin{tabular}{|c|c|c|c|c|}
\hline $\Delta t$ [months] & \multicolumn{2}{c|}{6} & \multicolumn{2}{c|}{3} \\
\hline
Model & $k^\alpha+f_0$ & $(k+k_0)^\alpha$ & $k^\alpha+f_0$ & $(k+k_0)^\alpha$
\\ \hline
$\alpha$ & $1.15 \pm 0.06$ & $1.24 \pm 0.14$ & $1.16 \pm 0.08$ & $1.24 \pm 
0.20$ \\
$f_0$ & $1.8 \pm 0.6$ & --- & $1.7 \pm 0.9$ & --- \\
$k_0$ & --- & $2.0 \pm 0.8$ & --- & $1.8 \pm 1.2$ \\ \hline
\end{tabular}
\caption{\label{preftable}%
Parameters obtained by fitting two alternative preferential-attachment kernels
to the citation data of category-4 patents granted in 1998.}
\end{table}

Conceptually, the two functional forms compared here differ in the way citation
growth is modeled for patents with small, especially vanishing, number of
citations. The underlying causes for determining the initial citation rate
(pure chance or some sort of quality/fitness) need to be investigated further.
The adherence to different functional forms of $f(k)$ revealed in the citation
rate for patents and scientific publications could be indicative of an
interesting distinction between growth mechanisms responsible for their
respective citation dynamics.

\subsection{Alternative forms of the ageing function}

Various forms of ageing functions have been considered in the context of
network-growth models~\cite{Wu2014}. Besides exponential ageing represented by
the functional form given in Eq.~(\ref{eq:ageing}) of the main text and deduced
also in the early economics literature on patent citations~\cite{jaffe1999},
expressions with an asymptotic power-law dependence $\lim_{t\to\infty} A(t) =
t^{-\beta}$ are quite common. In particular, previous studies of citation
dynamics~\cite{csardi2007,golosovsky2012,golosovsky2013} have fitted an ageing
function of the form
\begin{equation}\label{eq:powerAge}
A(t) = \frac{A_0^\prime}{\left( 1 + t/\tau^\prime \right)^\beta}
\end{equation}
to the data. Within our approach where patent-citation data are disambiguated
by technology category, citation inflation is accounted for, and the
application date of patents constitutes the time when incoming citations are
generated, the ageing function (\ref{eq:powerAge}) can quite accurately represent
the data, even at short times. See Fig.~\ref{fig:powerlaw}(a). However, as
illustrated Figs.~\ref{fig:powerlaw}(b) and \ref{fig:powerlaw}(c), the parameters
$\tau^\prime$ and $\beta$ extracted from such superficially accurate fits turn
out to have excessively large uncertainties and even show some residual $k$
dependence. Table~\ref{altparametertable} summarizes the consistency of this
situation across all technology categories. Thus while the introduction of an
additional parameter enables a good fit, no meaningful information can be
associated with the fitting parameters. The same type of problem occurs with
other three-parameter models such as a stretched-exponential expression and the
Weibull form used in Ref.~\cite{valverde2007}. In contrast to the explored
alternatives, fitting exponential ageing to the data at long times yields robust
and meaningful results that enable, e.g., comparisons between different
technology categories. These fits also reveal a systematic excess of citations at
short times that points to a different mechanism for knowledge flow from that
acting at longer times, which warrants further investigation. See
Fig.~\ref{fig:powerlaw}(a) and information given in the main text.

\begin{table*}
\begin{tabular}{|c|cccccc|}
\hline Category & 1 & 2 & 3 & 4 & 5 & 6 \\ \hline
$\beta$ & $1.4 \pm 0.5$\footnote{\centering{Parameter shows a significant
residual dependence on the cumulative number of citations.}} & $1.3 \pm
0.4^\mathrm{a}$ & $1.6 \pm 0.6$ & $1.4 \pm 0.3$ & $1.3 \pm 0.3$ & $1.3 \pm 0.3$
\\ $\tau^\prime$ [years] & $2.0 \pm 2.4^\mathrm{a}$ & $1.1 \pm 1.2^\mathrm{a}$
& $2.8 \pm 2.3$ & $1.9 \pm 1.2$ & $1.8 \pm 1.4$ & $2.2 \pm 2.1^\mathrm{a}$ \\
$A_0^\prime$ [years$^{-1}$] & $0.67\pm 0.06$ & $0.70\pm 0.02$ & $0.68\pm 0.03$
& $0.54\pm 0.04$ & $0.34\pm 0.07$ & $0.30\pm 0.05$ \\ \hline
\end{tabular}
\caption{\label{altparametertable}%
Parameters extracted from fits of patent-citation data to the power-law
expression for the ageing function given in Eq.~(\ref{eq:powerAge}). Values
obtained for $A_0^\prime$ are systematically larger than those for $A_0$ from
Table~\ref{parametertable} because the shifted-power-law fits include the data
at short times that deviate significantly from the exponential-ageing model.
[See Fig.~\ref{fig:powerlaw}(a).] Despite the apparent accuracy of
the fit that can be achieved by using a three-parameter model, excessively
large uncertainties of the extracted values for $\tau^\prime$ and $\beta$ in
all categories prevent its meaningful application.}
\end{table*}

The tendency to find power laws as ageing functions in previous work could be
explained by several mechanisms that conspire to fatten the tail of the
citation distributions. Most importantly, the ever-increasing overall number of
citing patents and scientific articles creates a citation inflation that, if
not accounted for, enhances citation rates at later times. Also, studies
employing the method of transitive reduction~\cite{clough2015} indicate that a
substantial number of citations in scientific articles are redundant for the
purpose of information flow, potentially being included mainly to honor
intellectual priority. This practice slows down obsolescence and establishes
a class of `immortal' papers~\cite{golosovsky2013}. No such effect exists for
patent citations, as the very small fraction of citations eliminated by
transitive reduction attests~\cite{clough2015}, thus exposing basic exponential
ageing at long times. Finally, the appearance of slower obsolescence could be
created through to the influence of autocorrelation effects.


\begin{thebibliography}{34}%
\makeatletter
\providecommand \@ifxundefined [1]{%
 \@ifx{#1\undefined}
}%
\providecommand \@ifnum [1]{%
 \ifnum #1\expandafter \@firstoftwo
 \else \expandafter \@secondoftwo
 \fi
}%
\providecommand \@ifx [1]{%
 \ifx #1\expandafter \@firstoftwo
 \else \expandafter \@secondoftwo
 \fi
}%
\providecommand \natexlab [1]{#1}%
\providecommand \enquote  [1]{``#1''}%
\providecommand \bibnamefont  [1]{#1}%
\providecommand \bibfnamefont [1]{#1}%
\providecommand \citenamefont [1]{#1}%
\providecommand \href@noop [0]{\@secondoftwo}%
\providecommand \href [0]{\begingroup \@sanitize@url \@href}%
\providecommand \@href[1]{\@@startlink{#1}\@@href}%
\providecommand \@@href[1]{\endgroup#1\@@endlink}%
\providecommand \@sanitize@url [0]{\catcode `\\12\catcode `\$12\catcode
  `\&12\catcode `\#12\catcode `\^12\catcode `\_12\catcode `\%12\relax}%
\providecommand \@@startlink[1]{}%
\providecommand \@@endlink[0]{}%
\providecommand \url  [0]{\begingroup\@sanitize@url \@url }%
\providecommand \@url [1]{\endgroup\@href {#1}{\urlprefix }}%
\providecommand \urlprefix  [0]{URL }%
\providecommand \Eprint [0]{\href }%
\providecommand \doibase [0]{http://dx.doi.org/}%
\providecommand \selectlanguage [0]{\@gobble}%
\providecommand \bibinfo  [0]{\@secondoftwo}%
\providecommand \bibfield  [0]{\@secondoftwo}%
\providecommand \translation [1]{[#1]}%
\providecommand \BibitemOpen [0]{}%
\providecommand \bibitemStop [0]{}%
\providecommand \bibitemNoStop [0]{.\EOS\space}%
\providecommand \EOS [0]{\spacefactor3000\relax}%
\providecommand \BibitemShut  [1]{\csname bibitem#1\endcsname}%
\let\auto@bib@innerbib\@empty
\bibitem [{\citenamefont {Redner}(1998)}]{redner1998}%
  \BibitemOpen
  \bibfield  {author} {\bibinfo {author} {\bibfnamefont {S.}~\bibnamefont
  {Redner}},\ }\href {\doibase 10.1007/s100510050359} {\bibfield  {journal}
  {\bibinfo  {journal} {Eur. Phys. J. B}\ }\textbf {\bibinfo {volume} {4}},\
  \bibinfo {pages} {131} (\bibinfo {year} {1998})}\BibitemShut {NoStop}%
\bibitem [{\citenamefont {B{\"o}rner}\ \emph {et~al.}(2004)\citenamefont
  {B{\"o}rner}, \citenamefont {Maru},\ and\ \citenamefont
  {Goldstone}}]{borner2004}%
  \BibitemOpen
  \bibfield  {author} {\bibinfo {author} {\bibfnamefont {K.}~\bibnamefont
  {B{\"o}rner}}, \bibinfo {author} {\bibfnamefont {J.~T.}\ \bibnamefont
  {Maru}}, \ and\ \bibinfo {author} {\bibfnamefont {R.~L.}\ \bibnamefont
  {Goldstone}},\ }\href {\doibase 10.1073/pnas.0307625100} {\bibfield
  {journal} {\bibinfo  {journal} {PNAS}\ }\textbf {\bibinfo {volume} {101}},\
  \bibinfo {pages} {5266} (\bibinfo {year} {2004})}\BibitemShut {NoStop}%
\bibitem [{\citenamefont {Redner}(2005)}]{redner2005}%
  \BibitemOpen
  \bibfield  {author} {\bibinfo {author} {\bibfnamefont {S.}~\bibnamefont
  {Redner}},\ }\href@noop {} {\bibfield  {journal} {\bibinfo  {journal} {Phys.
  Today}\ }\textbf {\bibinfo {volume} {58\rm{(6)}}},\ \bibinfo {pages} {49}
  (\bibinfo {year} {2005})}\BibitemShut {NoStop}%
\bibitem [{\citenamefont {Radicchi}\ \emph {et~al.}(2008)\citenamefont
  {Radicchi}, \citenamefont {Fortunato},\ and\ \citenamefont
  {Castellano}}]{radicchi2008}%
  \BibitemOpen
  \bibfield  {author} {\bibinfo {author} {\bibfnamefont {F.}~\bibnamefont
  {Radicchi}}, \bibinfo {author} {\bibfnamefont {S.}~\bibnamefont {Fortunato}},
  \ and\ \bibinfo {author} {\bibfnamefont {C.}~\bibnamefont {Castellano}},\
  }\href {\doibase 10.1073/pnas.0806977105} {\bibfield  {journal} {\bibinfo
  {journal} {PNAS}\ }\textbf {\bibinfo {volume} {105}},\ \bibinfo {pages}
  {17268} (\bibinfo {year} {2008})}\BibitemShut {NoStop}%
\bibitem [{\citenamefont {Golosovsky}\ and\ \citenamefont
  {Solomon}(2012)}]{golosovsky2012}%
  \BibitemOpen
  \bibfield  {author} {\bibinfo {author} {\bibfnamefont {M.}~\bibnamefont
  {Golosovsky}}\ and\ \bibinfo {author} {\bibfnamefont {S.}~\bibnamefont
  {Solomon}},\ }\href {\doibase 10.1103/PhysRevLett.109.098701} {\bibfield
  {journal} {\bibinfo  {journal} {Phys. Rev. Lett.}\ }\textbf {\bibinfo
  {volume} {109}},\ \bibinfo {pages} {098701} (\bibinfo {year}
  {2012})}\BibitemShut {NoStop}%
\bibitem [{\citenamefont {Golosovsky}\ and\ \citenamefont
  {Solomon}(2013)}]{golosovsky2013}%
  \BibitemOpen
  \bibfield  {author} {\bibinfo {author} {\bibfnamefont {M.}~\bibnamefont
  {Golosovsky}}\ and\ \bibinfo {author} {\bibfnamefont {S.}~\bibnamefont
  {Solomon}},\ }\href {\doibase 10.1007/s10955-013-0714-z} {\bibfield
  {journal} {\bibinfo  {journal} {J. Stat. Phys.}\ }\textbf {\bibinfo {volume}
  {151}},\ \bibinfo {pages} {340} (\bibinfo {year} {2013})}\BibitemShut
  {NoStop}%
\bibitem [{\citenamefont {Griliches}(1990)}]{Griliches1990}%
  \BibitemOpen
  \bibfield  {author} {\bibinfo {author} {\bibfnamefont {Z.}~\bibnamefont
  {Griliches}},\ }\href
  {http://EconPapers.repec.org/RePEc:aea:jeclit:v:28:y:1990:i:4:p:1661-1707}
  {\bibfield  {journal} {\bibinfo  {journal} {J. Econ. Lit.}\ }\textbf
  {\bibinfo {volume} {28}},\ \bibinfo {pages} {1661} (\bibinfo {year}
  {1990})}\BibitemShut {NoStop}%
\bibitem [{\citenamefont {Jaffe}\ and\ \citenamefont
  {Trajtenberg}(1999)}]{jaffe1999}%
  \BibitemOpen
  \bibfield  {author} {\bibinfo {author} {\bibfnamefont {A.~B.}\ \bibnamefont
  {Jaffe}}\ and\ \bibinfo {author} {\bibfnamefont {M.}~\bibnamefont
  {Trajtenberg}},\ }\href {\doibase 10.1080/10438599900000006} {\bibfield
  {journal} {\bibinfo  {journal} {Econ. Innov. New Techn.}\ }\textbf {\bibinfo
  {volume} {8}},\ \bibinfo {pages} {105} (\bibinfo {year} {1999})}\BibitemShut
  {NoStop}%
\bibitem [{\citenamefont {{von Wartburg}}\ \emph {et~al.}(2005)\citenamefont
  {{von Wartburg}}, \citenamefont {Teichert},\ and\ \citenamefont
  {Rost}}]{vonWartburg2005}%
  \BibitemOpen
  \bibfield  {author} {\bibinfo {author} {\bibfnamefont {I.}~\bibnamefont {{von
  Wartburg}}}, \bibinfo {author} {\bibfnamefont {T.}~\bibnamefont {Teichert}},
  \ and\ \bibinfo {author} {\bibfnamefont {K.}~\bibnamefont {Rost}},\ }\href
  {\doibase http://dx.doi.org/10.1016/j.respol.2005.08.001} {\bibfield
  {journal} {\bibinfo  {journal} {Res. Policy}\ }\textbf {\bibinfo {volume}
  {34}},\ \bibinfo {pages} {1591 } (\bibinfo {year} {2005})},\ \bibinfo {note}
  {{and citations therein.}}\BibitemShut {Stop}%
\bibitem [{\citenamefont {Cs{\'a}rdi}\ \emph {et~al.}(2007)\citenamefont
  {Cs{\'a}rdi}, \citenamefont {Strandburg}, \citenamefont {Zal{\'a}nyi},
  \citenamefont {Tobochnik},\ and\ \citenamefont {{\'E}rdi}}]{csardi2007}%
  \BibitemOpen
  \bibfield  {author} {\bibinfo {author} {\bibfnamefont {G.}~\bibnamefont
  {Cs{\'a}rdi}}, \bibinfo {author} {\bibfnamefont {K.~J.}\ \bibnamefont
  {Strandburg}}, \bibinfo {author} {\bibfnamefont {L.}~\bibnamefont
  {Zal{\'a}nyi}}, \bibinfo {author} {\bibfnamefont {J.}~\bibnamefont
  {Tobochnik}}, \ and\ \bibinfo {author} {\bibfnamefont {P.}~\bibnamefont
  {{\'E}rdi}},\ }\href {\doibase http://dx.doi.org/10.1016/j.physa.2006.08.022}
  {\bibfield  {journal} {\bibinfo  {journal} {Physica A}\ }\textbf {\bibinfo
  {volume} {374}},\ \bibinfo {pages} {783} (\bibinfo {year}
  {2007})}\BibitemShut {NoStop}%
\bibitem [{\citenamefont {Valverde}\ \emph {et~al.}(2007)\citenamefont
  {Valverde}, \citenamefont {Sol{\'e}}, \citenamefont {Bedau},\ and\
  \citenamefont {Packard}}]{valverde2007}%
  \BibitemOpen
  \bibfield  {author} {\bibinfo {author} {\bibfnamefont {S.}~\bibnamefont
  {Valverde}}, \bibinfo {author} {\bibfnamefont {R.~V.}\ \bibnamefont
  {Sol{\'e}}}, \bibinfo {author} {\bibfnamefont {M.~A.}\ \bibnamefont {Bedau}},
  \ and\ \bibinfo {author} {\bibfnamefont {N.}~\bibnamefont {Packard}},\ }\href
  {\doibase 10.1103/PhysRevE.76.056118} {\bibfield  {journal} {\bibinfo
  {journal} {Phys. Rev. E}\ }\textbf {\bibinfo {volume} {76}},\ \bibinfo
  {pages} {056118} (\bibinfo {year} {2007})}\BibitemShut {NoStop}%
\bibitem [{\citenamefont {Sheridan}\ \emph {et~al.}(2012)\citenamefont
  {Sheridan}, \citenamefont {Yagahara},\ and\ \citenamefont
  {Shimodaira}}]{sheridan2012}%
  \BibitemOpen
  \bibfield  {author} {\bibinfo {author} {\bibfnamefont {P.}~\bibnamefont
  {Sheridan}}, \bibinfo {author} {\bibfnamefont {Y.}~\bibnamefont {Yagahara}},
  \ and\ \bibinfo {author} {\bibfnamefont {H.}~\bibnamefont {Shimodaira}},\
  }\href {\doibase http://dx.doi.org/10.1016/j.physa.2012.05.041} {\bibfield
  {journal} {\bibinfo  {journal} {Physica A}\ }\textbf {\bibinfo {volume}
  {391}},\ \bibinfo {pages} {5031} (\bibinfo {year} {2012})}\BibitemShut
  {NoStop}%
\bibitem [{\citenamefont {{de Solla Price}}(1976)}]{sollaprice1976}%
  \BibitemOpen
  \bibfield  {author} {\bibinfo {author} {\bibfnamefont {D.}~\bibnamefont {{de
  Solla Price}}},\ }\href@noop {} {\bibfield  {journal} {\bibinfo  {journal}
  {J. Amer. Soc. Inform. Sci.}\ }\textbf {\bibinfo {volume} {27}},\ \bibinfo
  {pages} {292} (\bibinfo {year} {1976})}\BibitemShut {NoStop}%
\bibitem [{\citenamefont {Barab{\'a}si}\ and\ \citenamefont
  {Albert}(1999)}]{barabasi1999}%
  \BibitemOpen
  \bibfield  {author} {\bibinfo {author} {\bibfnamefont {A.-L.}\ \bibnamefont
  {Barab{\'a}si}}\ and\ \bibinfo {author} {\bibfnamefont {R.}~\bibnamefont
  {Albert}},\ }\href@noop {} {\bibfield  {journal} {\bibinfo  {journal}
  {Science}\ }\textbf {\bibinfo {volume} {286}},\ \bibinfo {pages} {509}
  (\bibinfo {year} {1999})}\BibitemShut {NoStop}%
\bibitem [{\citenamefont {Dorogovtsev}\ \emph {et~al.}(2000)\citenamefont
  {Dorogovtsev}, \citenamefont {Mendes},\ and\ \citenamefont
  {Samukhin}}]{dorogovt2000a}%
  \BibitemOpen
  \bibfield  {author} {\bibinfo {author} {\bibfnamefont {S.~N.}\ \bibnamefont
  {Dorogovtsev}}, \bibinfo {author} {\bibfnamefont {J.~F.~F.}\ \bibnamefont
  {Mendes}}, \ and\ \bibinfo {author} {\bibfnamefont {A.~N.}\ \bibnamefont
  {Samukhin}},\ }\href {\doibase 10.1103/PhysRevLett.85.4633} {\bibfield
  {journal} {\bibinfo  {journal} {Phys. Rev. Lett.}\ }\textbf {\bibinfo
  {volume} {85}},\ \bibinfo {pages} {4633} (\bibinfo {year}
  {2000})}\BibitemShut {NoStop}%
\bibitem [{\citenamefont {Krapivsky}\ and\ \citenamefont
  {Redner}(2001)}]{krapivsky2001}%
  \BibitemOpen
  \bibfield  {author} {\bibinfo {author} {\bibfnamefont {P.~L.}\ \bibnamefont
  {Krapivsky}}\ and\ \bibinfo {author} {\bibfnamefont {S.}~\bibnamefont
  {Redner}},\ }\href {\doibase 10.1103/PhysRevE.63.066123} {\bibfield
  {journal} {\bibinfo  {journal} {Phys. Rev. E}\ }\textbf {\bibinfo {volume}
  {63}},\ \bibinfo {pages} {066123} (\bibinfo {year} {2001})}\BibitemShut
  {NoStop}%
\bibitem [{\citenamefont {Albert}\ and\ \citenamefont
  {Barab{\'a}si}(2002)}]{albert2002}%
  \BibitemOpen
  \bibfield  {author} {\bibinfo {author} {\bibfnamefont {R.}~\bibnamefont
  {Albert}}\ and\ \bibinfo {author} {\bibfnamefont {A.-L.}\ \bibnamefont
  {Barab{\'a}si}},\ }\href {\doibase 10.1103/RevModPhys.74.47} {\bibfield
  {journal} {\bibinfo  {journal} {Rev. Mod. Phys.}\ }\textbf {\bibinfo {volume}
  {74}},\ \bibinfo {pages} {47} (\bibinfo {year} {2002})}\BibitemShut {NoStop}%
\bibitem [{\citenamefont {Newman}(2003)}]{newman2003}%
  \BibitemOpen
  \bibfield  {author} {\bibinfo {author} {\bibfnamefont {M.~E.~J.}\
  \bibnamefont {Newman}},\ }\href {\doibase 10.1137/S003614450342480}
  {\bibfield  {journal} {\bibinfo  {journal} {SIAM Review}\ }\textbf {\bibinfo
  {volume} {45}},\ \bibinfo {pages} {167} (\bibinfo {year} {2003})}\BibitemShut
  {NoStop}%
\bibitem [{Note1()}]{Note1}%
  \BibitemOpen
  \bibinfo {note} {The intrinsic citation rate $\lambda (t)$ is obtained from
  the bare total citation rate $\lambda _\protect \mathrm {tot}(t)$ by a
  rescaling to account for the extrinsic variation in citability due to the
  changing number $N(t)$ of patents that are generated at time $t$: $\lambda
  (t) = \lambda _\protect \mathrm {tot}(t)\protect \tmspace +\thinmuskip
  {.1667em} N(t_\protect \mathrm {a})/N(t)$. $t_\protect \mathrm {a} =
  0.5$~years in our analysis.}\BibitemShut {Stop}%
\bibitem [{\citenamefont {Dorogovtsev}\ and\ \citenamefont
  {Mendes}(2000)}]{dorogovt2000b}%
  \BibitemOpen
  \bibfield  {author} {\bibinfo {author} {\bibfnamefont {S.~N.}\ \bibnamefont
  {Dorogovtsev}}\ and\ \bibinfo {author} {\bibfnamefont {J.~F.~F.}\
  \bibnamefont {Mendes}},\ }\href {\doibase 10.1103/PhysRevE.62.1842}
  {\bibfield  {journal} {\bibinfo  {journal} {Phys. Rev. E}\ }\textbf {\bibinfo
  {volume} {62}},\ \bibinfo {pages} {1842} (\bibinfo {year}
  {2000})}\BibitemShut {NoStop}%
\bibitem [{\citenamefont {Zhu}\ \emph {et~al.}(2003)\citenamefont {Zhu},
  \citenamefont {Wang},\ and\ \citenamefont {Zhu}}]{zhu2003}%
  \BibitemOpen
  \bibfield  {author} {\bibinfo {author} {\bibfnamefont {H.}~\bibnamefont
  {Zhu}}, \bibinfo {author} {\bibfnamefont {X.}~\bibnamefont {Wang}}, \ and\
  \bibinfo {author} {\bibfnamefont {J.-Y.}\ \bibnamefont {Zhu}},\ }\href
  {\doibase 10.1103/PhysRevE.68.056121} {\bibfield  {journal} {\bibinfo
  {journal} {Phys. Rev. E}\ }\textbf {\bibinfo {volume} {68}},\ \bibinfo
  {pages} {056121} (\bibinfo {year} {2003})}\BibitemShut {NoStop}%
\bibitem [{\citenamefont {Medo}\ \emph {et~al.}(2011)\citenamefont {Medo},
  \citenamefont {Cimini},\ and\ \citenamefont {Gualdi}}]{Medo2011}%
  \BibitemOpen
  \bibfield  {author} {\bibinfo {author} {\bibfnamefont {M.}~\bibnamefont
  {Medo}}, \bibinfo {author} {\bibfnamefont {G.}~\bibnamefont {Cimini}}, \ and\
  \bibinfo {author} {\bibfnamefont {S.}~\bibnamefont {Gualdi}},\ }\href
  {\doibase 10.1103/PhysRevLett.107.238701} {\bibfield  {journal} {\bibinfo
  {journal} {Phys. Rev. Lett.}\ }\textbf {\bibinfo {volume} {107}},\ \bibinfo
  {pages} {238701} (\bibinfo {year} {2011})}\BibitemShut {NoStop}%
\bibitem [{\citenamefont {Wu}\ \emph {et~al.}(2014)\citenamefont {Wu},
  \citenamefont {Fu},\ and\ \citenamefont {Chiu}}]{Wu2014}%
  \BibitemOpen
  \bibfield  {author} {\bibinfo {author} {\bibfnamefont {Y.}~\bibnamefont
  {Wu}}, \bibinfo {author} {\bibfnamefont {T.~Z.}\ \bibnamefont {Fu}}, \ and\
  \bibinfo {author} {\bibfnamefont {D.~M.}\ \bibnamefont {Chiu}},\ }\href
  {\doibase http://dx.doi.org/10.1016/j.joi.2014.06.002} {\bibfield  {journal}
  {\bibinfo  {journal} {J. Informetrics}\ }\textbf {\bibinfo {volume} {8}},\
  \bibinfo {pages} {650 } (\bibinfo {year} {2014})}\BibitemShut {NoStop}%
\bibitem [{\citenamefont {Hall}\ \emph {et~al.}(2002)\citenamefont {Hall},
  \citenamefont {Jaffe},\ and\ \citenamefont {Trajtenberg}}]{hall2003}%
  \BibitemOpen
  \bibfield  {author} {\bibinfo {author} {\bibfnamefont {B.~H.}\ \bibnamefont
  {Hall}}, \bibinfo {author} {\bibfnamefont {A.~B.}\ \bibnamefont {Jaffe}}, \
  and\ \bibinfo {author} {\bibfnamefont {M.}~\bibnamefont {Trajtenberg}},\ }in\
  \href@noop {} {\emph {\bibinfo {booktitle} {Patents, Citations, and
  Innovations: A Window on the Knowledge Economy}}},\ \bibinfo {editor} {edited
  by\ \bibinfo {editor} {\bibfnamefont {A.~B.}\ \bibnamefont {Jaffe}}\ and\
  \bibinfo {editor} {\bibfnamefont {M.}~\bibnamefont {Trajtenberg}}}\ (\bibinfo
   {publisher} {MIT Press},\ \bibinfo {address} {Cambridge, MA},\ \bibinfo
  {year} {2002})\ p.\ \bibinfo {pages} {403}\BibitemShut {NoStop}%
\bibitem [{Note2()}]{Note2}%
  \BibitemOpen
  \bibinfo {note} {Three aspects -- disaggregation by technology,
  citation-inflation adjustment, and use of the time lag between granting of a
  cited patent and the \protect \emph {application date} of citing patents as
  the relevant time parameter -- differentiate our present approach from the
  one followed by previous studies of citation data~\cite
  {csardi2007,valverde2007, golosovsky2012}.}\BibitemShut {Stop}%
\bibitem [{Note3()}]{Note3}%
  \BibitemOpen
  \bibinfo {note} {More recent works~\cite {csardi2007,golosovsky2012}
  postulated ageing functions with asymptotic power-law behavior $A(t)\sim
  t^{-\beta }$ but concomitantly observed a marked increase of the
  preferential-attachment exponent $\alpha $ over time. Others \cite
  {valverde2007} presumed the ageing function to be of Weibull form but fixed
  $\alpha =1$ in their fits. As shown in Appendix~\ref {app:alternatives},
  exponential ageing best describes the intrinsic citation rate for patents in
  the long run.}\BibitemShut {Stop}%
\bibitem [{Note4()}]{Note4}%
  \BibitemOpen
  \bibinfo {note} {Solution of Eq.~(\ref {eq:zeroCit}) yields $n(0,t) = n_0(0)
  \left [ \gamma (t) \right ]^{f(0)}$. For given $n(k-1, t)$, Eq.~(\ref
  {eq:finiteCit}) is solved by the method of variation of constants, yielding
  $n(k, t) = n_0(k) \left [ \gamma (t) \right ]^{f(k)}+\left [ \gamma (t)
  \right ]^{f(k)} \DOTSI \intop \ilimits@ _0^t dt' \protect \tmspace
  +\thinmuskip {.1667em}\protect \tmspace +\thinmuskip {.1667em} \protect
  \mathaccentV {bar}016\lambda (k-1, t')\protect \tmspace +\thinmuskip
  {.1667em} \left [\gamma (t') \right ]^{-f(k)} \protect \tmspace +\thinmuskip
  {.1667em} n(k-1, t')$. The explicit result given in Eq.~(\ref
  {eq:citDistrib}) is obtained by induction.}\BibitemShut {Stop}%
\bibitem [{Note5()}]{Note5}%
  \BibitemOpen
  \bibinfo {note} {For $\alpha =1$, $f_0=1$, $n_0(k)=\delta _{k0}$ as assumed,
  e.g., in Refs.~\cite {Medo2011,Wu2014}, the result (\ref {eq:citDistrib})
  specializes to $n(k,t)=\gamma (t) \left [ 1 - \gamma (t)\right ]^k$, yielding
  $\Lambda (t)=A(t)/ \gamma (t)$ and $K(t)= \left [ \gamma (t)\right ]^{-1} -
  1$.}\BibitemShut {Stop}%
\bibitem [{\citenamefont {Jaffe}\ and\ \citenamefont
  {{de~Rassenfosse}}(2016)}]{jaffe2016}%
  \BibitemOpen
  \bibfield  {author} {\bibinfo {author} {\bibfnamefont {A.~B.}\ \bibnamefont
  {Jaffe}}\ and\ \bibinfo {author} {\bibfnamefont {G.}~\bibnamefont
  {{de~Rassenfosse}}},\ }\href {\doibase 10.3386/w21868} {\emph {\bibinfo
  {title} {Patent Citation Data in Social Science Research: Overview and Best
  Practices}}},\ \bibinfo {type} {Working Paper}\ \bibinfo {number} {21868}\
  (\bibinfo  {institution} {National Bureau of Economic Research},\ \bibinfo
  {year} {2016})\BibitemShut {NoStop}%
\bibitem [{Note6()}]{Note6}%
  \BibitemOpen
  \bibinfo {note} {TR removes links without disrupting `information flow', such
  that, when patent A cites patents B and C, and B also cites C, then the edge
  connecting A to C is removed, as the information still flows from C to A via
  B. This edge-removal process is argued to highlight the causal structure of a
  directed acyclic network, for which the resulting graph is unique. The
  authors of Ref.~\cite {clough2015} find that TR removes about 80\% of links
  in academic citation networks, but only 15\% in patent-citation
  networks.}\BibitemShut {Stop}%
\bibitem [{\citenamefont {Clough}\ \emph {et~al.}(2015)\citenamefont {Clough},
  \citenamefont {Gollings}, \citenamefont {Loach},\ and\ \citenamefont
  {Evans}}]{clough2015}%
  \BibitemOpen
  \bibfield  {author} {\bibinfo {author} {\bibfnamefont {J.~R.}\ \bibnamefont
  {Clough}}, \bibinfo {author} {\bibfnamefont {J.}~\bibnamefont {Gollings}},
  \bibinfo {author} {\bibfnamefont {T.~V.}\ \bibnamefont {Loach}}, \ and\
  \bibinfo {author} {\bibfnamefont {T.~S.}\ \bibnamefont {Evans}},\ }\href
  {\doibase 10.1093/comnet/cnu039} {\bibfield  {journal} {\bibinfo  {journal}
  {J. Complex Networks}\ }\textbf {\bibinfo {volume} {3}},\ \bibinfo {pages}
  {189} (\bibinfo {year} {2015})}\BibitemShut {NoStop}%
\bibitem [{Note7()}]{Note7}%
  \BibitemOpen
  \bibinfo {note} {This is illustrated by a large survey of inventors~\cite
  {jaffe2000} finding that almost half of the citations on their patents
  referenced inventions unknown to the inventors before the
  survey.}\BibitemShut {Stop}%
\bibitem [{\citenamefont {Cotropia}\ \emph {et~al.}(2013)\citenamefont
  {Cotropia}, \citenamefont {Lemley},\ and\ \citenamefont
  {Sampat}}]{cotropia2013}%
  \BibitemOpen
  \bibfield  {author} {\bibinfo {author} {\bibfnamefont {C.~A.}\ \bibnamefont
  {Cotropia}}, \bibinfo {author} {\bibfnamefont {M.~A.}\ \bibnamefont
  {Lemley}}, \ and\ \bibinfo {author} {\bibfnamefont {B.}~\bibnamefont
  {Sampat}},\ }\href {\doibase http://dx.doi.org/10.1016/j.respol.2013.01.003}
  {\bibfield  {journal} {\bibinfo  {journal} {Res. Policy}\ }\textbf {\bibinfo
  {volume} {42}},\ \bibinfo {pages} {844} (\bibinfo {year} {2013})}\BibitemShut
  {NoStop}%
\bibitem [{\citenamefont {Jaffe}\ \emph {et~al.}(2000)\citenamefont {Jaffe},
  \citenamefont {Trajtenberg},\ and\ \citenamefont {Fogarty}}]{jaffe2000}%
  \BibitemOpen
  \bibfield  {author} {\bibinfo {author} {\bibfnamefont {A.~B.}\ \bibnamefont
  {Jaffe}}, \bibinfo {author} {\bibfnamefont {M.}~\bibnamefont {Trajtenberg}},
  \ and\ \bibinfo {author} {\bibfnamefont {M.~S.}\ \bibnamefont {Fogarty}},\
  }\href {\doibase 10.3386/w7631} {\emph {\bibinfo {title} {The Meaning of
  Patent Citations: Report on the NBER/Case-Western Reserve Survey of
  Patentees}}},\ \bibinfo {type} {Working Paper}\ \bibinfo {number} {7631}\
  (\bibinfo  {institution} {National Bureau of Economic Research},\ \bibinfo
  {year} {2000})\BibitemShut {NoStop}%
\end{thebibliography}
%

\end{document}